\documentclass[article,3p,twocolumn]{elsarticle}

\usepackage{lipsum}
\usepackage{widetext}

\usepackage{dcolumn}
\usepackage{epsfig}

\newcommand{\tabref}[1]{Table \ref{#1}}
\newcommand{\figref}[1]{Fig. \ref{#1}}

\newcommand{\fs}{\begin{equation} \label}
\newcommand{\fe}{\end{equation}}

\begin{document}

\begin{abstract}
Calculating interactions or correlations between pairs of particles is
typically the most time-consuming task in particle simulation
or correlation analysis. Straightforward implementations using a double
loop over particle pairs have traditionally worked well, especially since compilers usually
do a good job of unrolling the inner loop. In order to reach high performance on
modern CPU and accelerator architectures, single-instruction multiple-data (SIMD) parallelization has become essential.
Avoiding memory bottlenecks is also increasingly important and requires
reducing the ratio of memory to arithmetic operations.
Moreover, when pairs only interact within a certain cut-off distance,
good SIMD utilization can only be achieved by reordering input and output data,
which quickly becomes a limiting factor.
Here we present an algorithm for SIMD parallelization
based on grouping a fixed number of particles, e.g. 2, 4, or 8, into spatial clusters.
Calculating all interactions between particles in a pair of such clusters improves
data reuse compared to the traditional scheme and results in a more efficient SIMD parallelization.
Adjusting the cluster size allows the algorithm to map to SIMD units of various widths. This flexibility not only enables fast and efficient implementation on current CPUs and accelerator architectures like GPUs or Intel MIC, but it also makes the algorithm future-proof.
We present the algorithm with an application to molecular dynamics simulations,
where we can also make use of the effective buffering the method introduces.

\end{abstract}

\title{A flexible algorithm for calculating pair interactions on SIMD architectures}
\author{Szil{\'a}rd P{\'a}ll \texttt{pszilard@kth.se}}
\author{Berk Hess \texttt{hess@kth.se}}
\address{Theoretical and Computational Biophysics, Department of Theoretical Physics and Swedish e-Science Research Center, KTH Royal Institute of Technology, 10691 Stockholm, Sweden}
\address{Science for Life Laboratory, Stockholm and Uppsala, 17121 Stockholm, Sweden}

\maketitle

\begin{widetext}
{\bf NOTICE:} this is the author’s version of a work that was accepted for publication in Computer Physics Communications. Changes resulting from the publishing process, such as peer review, editing, corrections, structural formatting, and other quality control mechanisms may not be reflected in this document. Changes may have been made to this work since it was submitted for publication.
\end{widetext}

\section*{Keywords}
Pair interactions, SIMD, GPU, Molecular Dynamics, Verlet list

\section{Introduction}
In most particle simulations, more than half of the computational time
is spent in calculating pair interactions with limited spatial range.
When long-range interactions are present, such as electrostatics,
the long-range part is usually calculated on a mesh.
Certain types of analysis, such as determining particle pair correlation
functions, also involve evaluating pair interactions with limited range.

Many codes that compute these kind of interactions employ CPU algorithms consisting of
a simple double loop to iterate through a list of particle pairs.
This na{\"i}ve approach has a quadratic computational complexity which makes it
prohibitively expensive already for moderate numbers of particles.
However, by exploiting the limited interaction range imposed by the typically spherical cut-off,
the computational cost can be reduced to linear.
This is achieved by reducing the number of neighboring particles that need to be
considered. To do so the Verlet list \cite{Verlet1967} and the linked cell \cite{Hockney74}
algorithms as well as the combination of the two are widely used.
In particular, in molecular dynamics (MD) simulation codes \cite{Plimpton1995,Phillips2005,Anderson2008,Hess2008b,Brown2011}
these algorithms are most commonly employed. Although these algorithms suffer from
limitations on modern SIMD architectures \cite{Pennycook2012,Pennycook2013}, there have been
only a few attempts to overcome them, most of them specific to GPUs \cite{Meel2008,Friedrichs2009}
without achieving generality.

Before the advent of CPU SIMD units, the performance of the simple double loop over the neighbor list was quite good
as the compiler can usually unroll the inner loop. Because the speed
of the main memory has not kept up with the processor speed, caching
became more important. In calculating pair interactions this means that the location
of particles in memory should correlate with their spatial location
to increase cache hits. Several publications have dealt with this issue \cite{Yao2004,Meloni2007,Hess2008b}.
However, as the width of the SIMD units increases, reordering or shuffling
the input and output data for convenient access in the SIMD units becomes
a severe bottleneck. When calculating
pair interactions between all particle pairs in the system, a perfectly
linear memory access pattern can be used that avoids shuffling.
However, when a cut-off is used, a significant part of the particle neighbor list will

not be ordered sequentially. The relative cost of shuffling depends
on the cost of calculating a single pair interaction and on the SIMD width.
In molecular dynamics simulations particles usually interact via
a Lennard-Jones (LJ) and a Coulomb potential. When the popular particle-mesh
Ewald (PME) electrostatics method \cite{Essmann95} is used, a complementary
error function must be calculated.
Pennycook et al. \cite{Pennycook2013} provide a detailed analysis of the shuffling 
(also called gather-scatter) and their impact on performance with only LJ interactions considered.
In their work, with 8-way SIMD reordering instructions represent a third of the total,
with 16-way SIMD the ratio is more than a half. In practice, the performance is affected
even more. Since shuffling introduces more data dependencies between
instructions, reducing the instructions available for scheduling
will result in low instructions per cycle (IPC). We will show that even when
calculating LJ and PME interactions, the shuffling ends up taking
more than half of the time with 4-way SIMD.

On GPUs, shuffling data is typically not required as the execution model allows hardware
threads to access data from different memory locations. However, loading particle data
requires scattered memory access which will waste GPU memory bandwidth as well as
cycles (due to instruction replay) and will render a standard implementation
memory bound. Moreover, the throughput-oriented GPU architecture requires
high level of parallelism and is sensitive to memory access patterns.
In order to target GPUs, some codes combine the traditional algorithms
with data regularization techniques \cite{Stone2007,Meel2008},
but such approaches can still lead to inefficient execution.
Recasting the algorithms to a more regular data access has been shown to result in higher
IPC on GPUs, but not without additional trade-offs \cite{Eastman2009}. Although on CPUs the relative memory bandwidth is higher,
the data dependencies can still cause bottlenecks in SIMD-optimized algorithms.

The main issues faced when considering data parallelization in traditional
particle-pair based neighbor-lists schemes are the irregular sizes and non-contiguous
nature of the neighbor lists of each particle. We propose to address both of
these issues by considering pair-interactions between clusters of particles
of fixed size, similar to the work of Friedrichs et al. \cite{Friedrichs2009}.
However, important distinguishing features of our algorithm are high parallel work-efficiency
and the inherent flexibility which enables tuning for the SIMD width and other specifics of the hardware.
By changing the size of the clusters, our algorithm can be
adapted to SIMD units of different widths. Adjusting the cluster size also
allows tuning the number of operations ``in flight'' as well as the
ratio of arithmetic to memory operations. This flexibility, together
with the high ratio of arithmetic to load/store operations, ensures
that the algorithm can reach high performance on current, as well as future
CPU and GPU hardware. It is also well suited to more exotic hardware
such as FPGAs, but as the implementation is still ongoing, result will be reported in the future.
In case of CPUs, the additional major advantage is that, by matching the cluster
size to the SIMD width, no shuffle operations are required at all. This not only
improves performance by at least a factor of 2, but also makes the code much
easier to write and read. 
There is a price to pay for the improvements as 
the cluster pairs will contain particle pairs in addition to the ones in the original interaction sphere. This results in extra interactions calculated between particles otherwise not within range, which we know will evaluate to zero. As we will show later,
although this does lead to reduction in algorithmic work-efficiency, the performance gain still outweighs the extra cost.

We would like to note that the algorithm operates on the lowest level
of the interaction calculation and any optimization available in the
literature can be applied. For MD, we use it together with a Verlet buffer.
Furthermore, all parallelization strategies developed for traditional
algorithms can be used with little or no modification.

We have designed and implemented non-bonded pair interaction kernels
for x86 SSE2, SSE4.1, AVX and AVX+FMA (AMD Bulldozer) SIMD architectures,
as well as NVIDIA GPUs.
The kernels utilize LJ interactions and monopole-monopole
electrostatic interactions of general form. We implemented analytical
electrostatics kernels for reaction-field (RF) and PME, as well as
tabulated electrostatic potentials. We plan to support sphero-symmetric
potential of arbitrary shape through tabulated interactions.
While the required additional table lookups per pair will lower the efficiency of the kernels
on current CPUs, on GPUs and with AVX2 (which will support table lookups)
performance should be good.

The algorithms described here have been implemented in the GROMACS
molecular simulation package \cite{Hess2008b,Pronk2013a} and are available in the
official version 4.6 release, combined with hybrid MPI+OpenMP parallelization.
The source code can be obtained under
the LGPLv2 license from {\tt http://www.gromacs.org}.
Note that the CPU kernels in GROMACS 4.6 have an additional optimization,
not discussed in this paper, for systems where less than half of the particles
have LJ interactions. For water this improves kernel performance by up to 10\%.

\section{The algorithm}
We are looking for an algorithm that can execute single instructions
on multiple data (SIMD), while not being limited by loading and storing data
from and to (cache-)memory.
The standard implementation of the Verlet-list algorithm
loads a particle and calculates pair interactions by looping over its neighbors.
Thus a single pair interaction is calculated for each particle load and store.
The relatively cheap interactions in MD simulations render this algorithm effectively memory bound. To remedy this, our algorithm loads a cluster of $M$ particles and calculate $M$ interactions for each neighbor loaded. This increases the data reuse by a factor of $M$.
The loop over neighboring particles is replaced by a loop over clusters consisting of $N$ particles. The values of $M$ and $N$ will be tuned for the SIMD hardware.
The standard implementation of the Verlet-list algorithm can be seen as a special case of this cluster algorithm where $M$=1 and $N$=1.

\subsection{Limitations of the standard implementation}
In general, the easiest way to achieve SIMD parallelization is to let the compiler
vectorize loops, possibly with the help of the programmer aided by feedback from the compiler.
At a first glance this might seem to be a good strategy since a particle usually
has hundreds of neighbors which leads to long vectorizable loops.
For efficient loading, the order of particles in memory needs to
be strongly correlated with spatial ordering to increase cache hits.
Ideally, sequential particles would be loaded in groups of size equal to
the SIMD width, but this not compatible with a spherical interaction volume.
Even when particles can be loaded in groups,
vectorizing the inner-loop will only give a small speed-up
on wider SIMD units, as memory operations and data shuffling can take more time
than the actual calculation.
For LJ only with fixed parameters on AVX 8-way SIMD, memory and shuffling
operations account for 32 of the 70 operations \cite{Pennycook2013};
with parameter loading, the ratio increases beyond 50\%.
When calculating all interactions of neighbors
with one particle, we need to load 3 coordinate components, 3 parameters, as well as
load and store 3 force components for each neighbor. In theory, on current CPUs
this should not lead to a memory-bound algorithm, but in
practice performance will be far from peak due to limitations
on the instruction scheduling. The coordinates are loaded per particle
as triplets of $x$, $y$, $z$ requiring data-shuffling. The wider the CPU SIMD unit is, the more
data shuffling is required and the longer the dependency chain gets between
loading data, computation and storing forces. Hence, for efficient SIMD calculations
it is very advantageous to use packed sequences of coordinates,
e.g $xxxx$, $yyyy$ and $zzzz$ with 4-way SIMD.
On GPUs, such packing is not needed as vector types are supported, but a much higher arithmetic
to memory operation ratio is required to achieve peak performance.
Constructing the neighbor list, also called pair list, is a similar operation, but with less arithmetic,
which makes it even more memory intensive. Although the pair list is usually not
reconstructed every step, it involves looping over more pairs than the
non-bonded kernel processes, so this can become a limiting factor.
The only way to hide the latency of memory operations\footnote{
The technique of overlapping multiple operations (typically arithmetic
with memory) in order to avoid the memory latency bottleneck is called
latency ``hiding.''
The amount of latency hiding possible depends on specifics of the architecture like
register set size, cache size and behavior, instruction scheduling, as well as compiler optimizations;
the mix and order of instructions used are the only factors which the programmer
can directly influence without writing raw assembly code.}
is to perform more calculation per load/store operation. At first sight this might seem impossible,
but this can actually be achieved with a simple scheme.

\subsection{The $M \times N$ algorithm}
The basic idea behind our work is to spatially cluster particles in groups of fixed size
and use such a cluster as the computational unit of our algorithm. These groups can then
be mapped directly to the SIMD hardware units, which have a fixed width.
Given a 4-way SIMD unit, we can spatially cluster particles in groups of 4.

\begin{figure}
\centerline{\includegraphics[width=8cm]{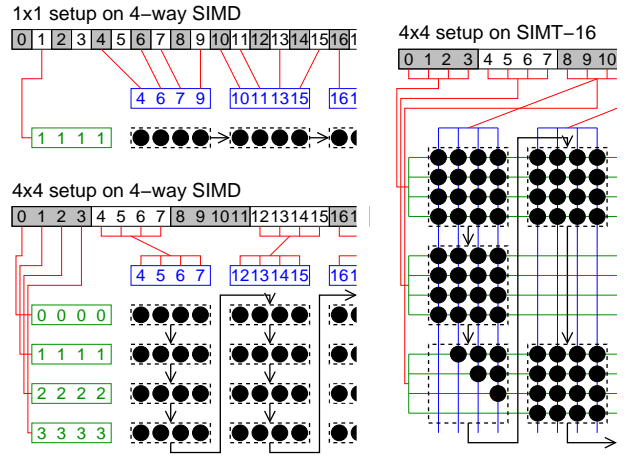}}
\caption{Illustration of the 1$\times$1 and the 4$\times$4 setups with 4-way SIMD and 16-way SIMT. All numbers are particle indices, each black dot represents an interaction calculation and the arrows indicate the computational flow. The SIMD registers for $i$- and $j$-particles are shown in green and blue, respectively. The 4$\times$4 setup calculates 4 times as many interactions per particle load/store and requires fewer memory operations (shown in red). Unlike the 1$\times$1 setup, the 4$\times$4 setup does not require data shuffling in registers.
\label{simd}}
\end{figure}

We can load a cluster of 4, so called, $i$-particles in SIMD registers and then loop
over the neighboring clusters of 4, so called, $j$-particles (see \figref{simd}). With this $M \times N$ = 4$\times$4 setup, we compute
16 pair interactions while only performing memory load and store operations
for 4 $j$-particles. After having looped over all neighboring $j$-clusters
of an $i$-cluster, usually a few hundred,
we also have to do memory operations for the $i$-particles,
but the cost of this is negligible. In this example the memory bandwidth
is reduced by a factor of 4, but more importantly, as we always access
particles in cluster of size 4, we can organize all data packed in groups
of 4. This eliminates the need for data shuffling which is the main
performance bottleneck of the standard way of calculating non-bonded
interactions on SIMD units.
This is the simplest version of the algorithm. The same 4$\times$4 clusters can also be processed on 8-way SIMD hardware. Then two i-clusters are loaded in one SIMD register and each j-cluster is duplicated in one SIMD register. This setup halves the number of arithmetic operations and adds a few shuffle operations.
In CUDA, memory access is more flexible. Hardware threads on NVIDIA GPUs are organized in ``warps''. On current GPUs, each warp consists of 32 threads which execute the same instruction every cycle.
This results in a SIMD-like execution model called single instruction multiple threads (SIMT).
Unlike the SIMD model which requires explicit programming for the SIMD width, the SIMT architecture allows thread-level parallel programming, and the warp-based lockstep execution model needs to be considered only for performance. This enables more flexible memory operations (different addresses in different threads) and divergence among threads in a warp.
The SIMT model allows spreading out all particle pairs in an 8$\times$4 cluster pair over the 32 threads of a warp, thus processing one particle pair on each thread. We illustrate this in \figref{simd}, for the sake of example using 16-way SIMT.

We will now describe in detail how the algorithm works, starting
with building the cluster pair list.

\subsection{Pair list construction}
The algorithmic unit of particle data representation is a cluster rather than a single particle.
Beside this minor, but important difference, the overall algorithm closely follows the
standard Verlet or neighbor list setup. Hence, in the following, unless explicitly stated, pair list will refer to a list of cluster pairs. Note that the cluster pair list this work uses as data representation does not define a strict particle-particle in-range   relationship because, as we will show later, the list by design includes particles not in-range. Moreover, the presented algorithms use Newton's third law to calculate pair interactions, hence the pair list contains each pair only once, not twice. Since for each particle there is no explicit list of all particles in its neighborhood, we prefer the term ``pair list'' to the term ``neighbor list''.

We construct a pair list using a Verlet buffer (also called ``skin'') which is essentially an extension of the cut-off distance to account for particle movement allowing the list to be retained for a number of steps \cite{Verlet1967}.
The exact number depends on the relative cost of the list construction and
the dependence of the buffer size on the lifetime of the list.
Pair interactions are then determined for the fixed list of particle pairs
defined by pairs of clusters.
In the most general case, we need to generate a pair list of clusters
of size $M$ particles versus clusters of size $N$.
In the simplest setup, the SIMD width will be
equal to $N$, but a width of $p\,N$, where $M$ is divisible by $p$,
will also work. On a GPU the best performance will be achieved when
matching $M \times N$ to the width of the SIMT execution model, i.e. 32 for CUDA.

\begin{figure}
\centerline{\includegraphics[width=6cm]{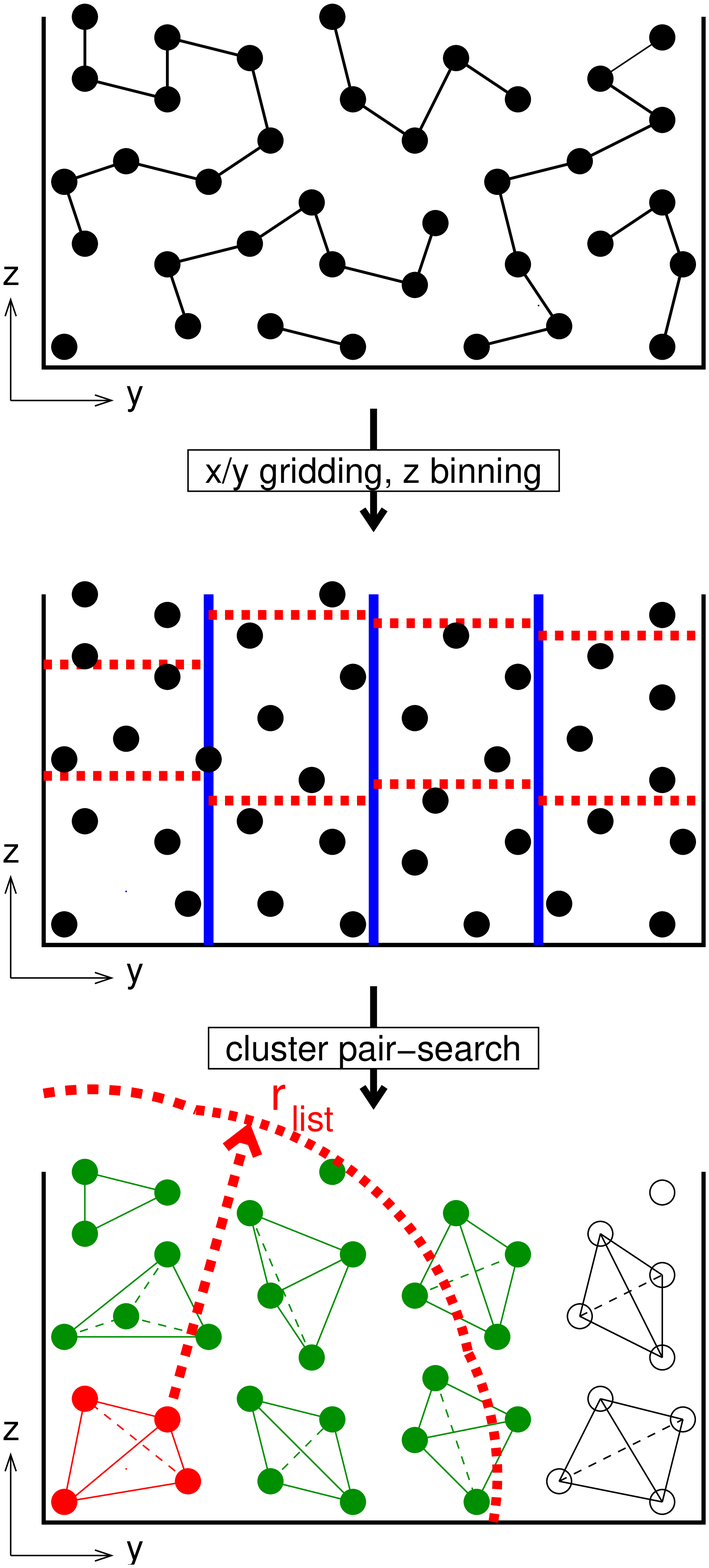}}
\caption{Illustration of the cluster pair-search algorithm for clusters of 4 particles. The bottom figure shows the $j$-cluster list in green for the red $i$-cluster. \label{clustering}}
\end{figure}

First we need to group the particles into clusters of fixed size.
To minimize the number of additional particle pairs in the pair list,
the clusters need to be as compact as possible.
A simple and efficient way of generating compact, fixed-size clusters
is spatial gridding in two dimensions and spatial binning
in the third dimension, see \figref{clustering}.
First we construct a rectangular grid along x and y with a
grid spacing of $(\max(M,N)/\rho)^{1/3}$, where $\rho$ is the particle density.
Then we sort the particles into columns of this grid.
For each column we sort the particles on z-coordinate and as a result
we get the spatial clusters as consecutive groups of $M$ or $N$ particles.
Because the number of particles in a column is
typically not a multiple of $M$, we add dummy particles to the last cluster when needed.
The fraction of dummy particles is $1/2\left( \max\left( M,N \right) \right)^{1/3}/\#\mathrm{particles}$;
with 10000 particles and clusters of size 8 this gives 4\% dummy particles
in the CPU algorithm. For the GPU we use a hierarchical cluster setup. As we can store
8 i-clusters in shared memory\footnote{Shared memory on NVIDIA GPUs is a fast on-chip memory,
essentially a programmable L1 cache.},
we group 2$\times$2$\times$2=8 clusters
of size 8 together. This reduces the number of dummy particles to 1\% with 10000 particles.
All these operations can be done efficiently in linear time.
The next step is calculating bounding boxes for each cluster,
this can be done using SIMD instructions, as the number of particles
in a cluster is constant.
In the case of $M \neq N$, adjacent pairs of bounding boxes are combined
to generate clusters of double the number of particles.
A pair list can then be constructed by checking distances
between the bounding boxes. This is very efficient, as it requires
one bounding box-pair distance check for $M\times N$ particle pairs.
However, this results in more cluster or particle pairs than strictly necessary,
as bounding boxes might be within range while none of the particle pairs
falls within range.
To avoid this overhead we prune pairs of clusters at distances close to the cut-off
using a particle-pair distance criterion.
For the GPU implementation, the pair list construction is performed on the CPU,
but the pruning is done on the GPU where this can be done more efficiently.
Periodic boundary conditions can be implemented in a simple and efficient
fashion by moving the $i$-clusters by the required periodic image shifts
and storing these shifts in the cluster pair list for use during the pair
interaction calculation.

\begin{figure*}
{\small
\begin{verbatim}
/* BB = cluster bounding box */
for each i-cluster ci
  determine grid range in x within rlist of BB[ci]
  for each grid cell gx in range
    determine grid range in y within rlist of BB[ci]
    for each grid cell gy in range
      determine j-clusters at gx,gy within rlist of BB[ci]
      for each j-cluster cj in range
        if (BBdistance(ci,cj) < rlist)
          if (BBdistance(ci,cj) < rBB or
              atomdistance(ci,cj) < rlist)
            put cj in cjlist[ci]
  set forcefield exclusion masks in cjlist[ci]
\end{verbatim}
}
\caption{Pseudocode for the cluster-pair search algorithm.
We avoid the expensive, up to $M \times N$ cost, atom-pair distance calculation
when the bounding box distance is shorter then {\tt rBB}.
\label{code_search}}
\end{figure*}

At this point we would like to note that the cluster pair list, being simply a Verlet list of particle clusters, can be seen as a generalized version of the classical neighbor list. Consequently, the neighbor list corresponds to the special case of cluster size $M = N = 1$ and in the following we will refer to it as the 1$\times$1 scheme.

\begin{figure}
\centerline{\includegraphics[width=7.5cm]{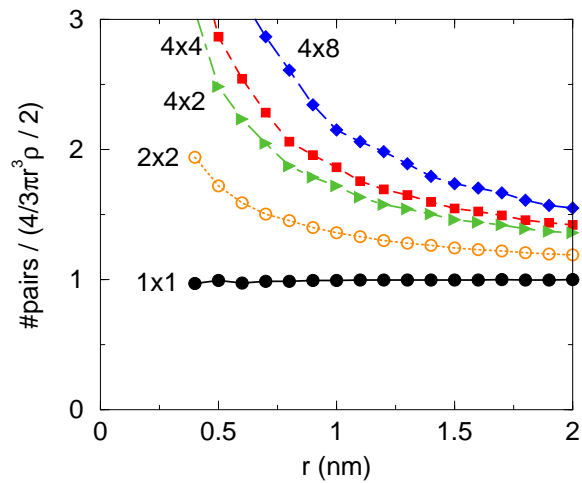}}
\caption{Number of pairs in different $M \times N$ pair lists normalized by the average number of pairs in a sphere of radius $r$, as a function of the pair list radius $r$, for a 3-site water model, number density $\rho$=100 nm$^{-3}$.
\label{zeros}}
\end{figure}

The cluster-based pair list contains inherently more particle pairs than the ones within
the cut-off radius, the number of pairs for different $M \times N$ is shown on \figref{zeros}.
The fraction of extra interactions increases rapidly with the cluster size and
decreases rapidly with the cut-off radius. However, the increase in the efficiency of the presented algorithm should outweigh the cost of calculating these extra interactions.
As can be seen in \figref{zeros}, when it comes to the number of extra pairs, lists with $M \neq N$ are less favorable than $N = M$ which results in clusters with close to cubic shape. This shape minimizes the number intersecting cut-off spheres, resulting in a more compact list.

\subsection{Exclusion treatment}
Having constructed the pair list,
for each cluster we now have the lists of all clusters in range.
We still need to take care of particle pairs that need to be excluded. There are
three types of exclusions. Two of those occur within cluster self-pairs.
Here particle pairs occur twice, whereas we should only calculate them once
and there are self interactions.
We want to calculate each pair interaction only once and skip the self
interactions. These two types of exclusions are handled in the pair interaction kernel.
Additionally, there can be exclusions defined by the force field. Normal LJ
and electrostatic interactions should not be calculated for such excluded
pairs of particles, whereas the RF or PME correction should still be applied.
To treat these exclusions in the non-bonded kernels, we encode them in a bitmask
stored per cluster pair in the pair list. This compact representation saves memory
and also allows for easy and fast decoding of exclusions using bitwise operations.
On CPUs we sort the pair list according to the presence of exclusions so we only
need to mask exclusions when really needed. This improves performance by 15\%.

\subsection{Computational cost}
\begin{figure}
\centerline{\includegraphics[width=7.5cm]{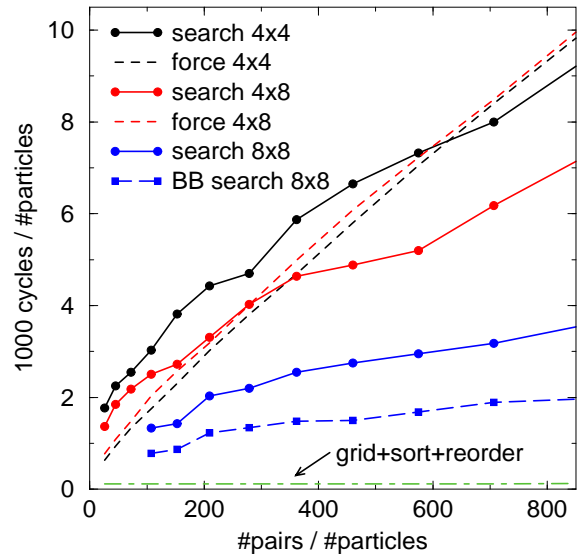}}
\caption{Computational cost in CPU cycles as a function of the number of pairs per particle for the search and LJ+PME force calculation for different $M \times N$, for 8$\times$8 the search cost is also shown for checking bounding box distances only. The pair count is within the spherical volume, not including the extra pairs due to the irregular cluster-pair volume. Note that the wiggles on the curves for searching are caused by jumps in the number of grid cells fitting in a cut-off sphere. All timings were done on an Intel Sandy Bridge CPU with single precision 256-bit AVX kernels using a single thread.
\label{cycles_npair}}
\end{figure}

The total computational cost of the pair list construction is proportional to the number of particles.
To understand how the total cost scales with different parameters, it is worth
looking into the details of the different tasks involved. Computational cost
in terms of cycles as a function of pairs per particle is shown in \figref{cycles_npair}.
For the implementation of the sorting, after the gridding, we assume
that the particle distribution is homogeneous on longer length scales, but other suitable sorting techniques can be applied in the inhomogeneous case.
Then a simple pigeonhole sorting is used, which scales linearly and provides good performance.
The cost of the pair search is not proportional to the number of pairs,
as only the boundary of the interaction sphere needs to be determined.
This cost is high when the number of pairs is small compared to $M$ and $N$
and it is proportional to the radius squared for large radius.
When the radius is large, the cost of search decreases proportionally with $M \times N$.
This makes the search far more efficient than a particle-pair based search.
Another implication of the much lower number of (now cluster-) pairs,
is that advanced search algorithms, such as interaction sorting  \cite{Gonnet2007, Welling2011},
will not help and a simple search algorithm performs well. Finally,
the cost of interaction calculation is proportional to the total number of pair
interactions calculated.
But the overhead of pairs beyond the cut-off distance decreases with
increasing number of pairs.
The cost of the search and LJ+PME force calculation on CPUs are similar,
as can been seen in \figref{cycles_npair}; RF kernels are about twice as fast,
which makes the search relatively more expensive.
The optimal balance between search cost and extra cost due to the Verlet buffer
is usually achieved with a pair list update interval of 10 and 20 when only using a CPU.
On the GPU, the interaction throughput is much higher which makes the CPU search
relatively more expensive. As mentioned, before, we do most of the cluster-pair
pruning on the GPU, which reduces the CPU search cost significantly, as can been
seen in \figref{cycles_npair}. Depending on the speed of the CPU versus the GPU, and especially the number
of cores in each, the optimal pair list update interval is between 10 and 50.
Furthermore, as the search algorithm maps well to GPUs \cite{Brown2011}, we plan to port it in the near future.

When the pair list needs to be updated for every interaction calculation,
the particle-pair distance based pruning should be skipped and replaced
by a conditional in the interaction kernels. With very cheap interactions, such
as for a pair correlation function calculation, no conditional should be used at all.

\section{Pair interaction kernels}
As in molecular simulations usually more than half of the computational time 
is spent in the calculating non-bonded pair interactions, it is well worth
carefully optimizing these kernels. We now have a list of cluster pairs of
$M$ versus $N$ particles. As can be seen in \figref{code_kernel}, writing a SIMD
kernel for this setup is rather straightforward.
However, achieving optimal performance is not trivial.
It is often very hard to judge how close the kernel performance is to the maximum achievable performance, as it depends both on hardware characteristics, mainly
the type of SIMD unit and the performance of the cache system and load/store units,
as well as on software characteristics, mainly the compiler(s) used.

\subsection{Kernel implementation}

\begin{figure}
{\small
\begin{verbatim}
for each ci cluster
  load M coords+params for ci

  for each cj cluster
    load N coords+params for cj
    /* These loops are unrolled using SIMD */
    for j=0 to M
      for i=0 to N
        calculate interaction ci*M+i with cj*N+j
    store N cj-forces
  store M ci-forces
\end{verbatim}
}
\caption{Pseudocode for the cluster-pair interaction kernel on CPUs.
\label{code_kernel}}
\end{figure}

For CPUs we chose to write the kernels in C with extensive use
of SSE and AVX SIMD-intrinsics\footnote{Intrinsic functions, also called
compiler intrinsics or builtins, are handled by the compiler in a
special manner, usually by replacing their call at compile time with a sequence of instructions.
SIMD intrinsics map to one or more assembly instructions
with additional optimization applied that integrate the inserted code.
Hence, with modern compilers, SIMD intrinsics achieve better performance than
equivalent inline assembly.}
as the current GNU and Intel compilers
do a good job at optimizing such code and typically achieve better performance
across multiple architectures than equivalent hand-written assembly. For GPUs we
chose to concentrate on the NVIDIA CUDA programming model as the 
available development tools are more mature and provide higher performance
than that of the alternatives.
There are two main factors that affect kernel performance and require special attention.
One is the choice of $M$ and $N$, the other is the treatment of the exclusion
and cut-off checks. For the latter the options are using conditionals,
which should be avoided on CPUs, or masking interactions using bitmasks.
Masking is usually more efficient than conditionals. On CPUs SIMD bitwise {\tt AND}
operations are used for masking, whereas on GPUs we simply multiply
by 0 or 1 and use a conditional for the cut-off check. Using a conditional can
reduce the number of instructions issued when all pairs stored in a SIMD
register are beyond the cut-off distance. On the CPU this should only be
used when all $M \times N$ pairs are beyond the cut-off, as otherwise
the force reduction cost increases. This only improves the performance
when an overly long pair list buffer is used, so we only use a conditional
for the 1$\times$1 kernels where it helps in most cases.
With the latest CPU compilers, not much code optimization is
required, as long as the fastest possible intrinsic is used for
the respective instruction set, e.g. SSE2, SSE4.1 or AVX.
In CUDA optimization is less straightforward; as the architecture is changing rapidly,
compilers and drivers are less mature. Additionally, GPUs are massively parallel processors
with more simple cores than CPUs, which puts more burden on the programmer and
compiler to pick the right optimization which might not even carry across hardware generations.

The main goal is to keep the computational units as busy as possible
by avoiding stalls due to dependencies on memory operations or instruction latencies.
The ratio of compute to memory operations scales with $M$, as for each
loaded $j$-particle, $M$ interactions with $M$ $i$-particles are calculated
in a single inner-loop iteration.
On the CPU it turns out that the best performance is achieved for $M$=4.
While using 2 $i$-particles is also possible, with 4 there seem to be enough
arithmetic instructions fed to the scheduler to hide most memory operations.
Using $M > 4$ will lead to a marginally higher IPC and flop rate, at the expense
of calculating many more zeros.
The choice of $N$ depends on the the SIMD width. Here, for CPUs, we only consider value of $N$ equal to the full or half SIMD width.
Fitting two $j$-clusters in a SIMD-width
will simply halve the number of arithmetic operations and add a few shuffle
operations.
The instruction count is largely independent of $N$. 
The only exception is the table lookup for tabulated electrostatics,
the number of which scales with both $M$ and $N$.
The precision, either single or double, also doesn't affect the instruction
count, except for the inverse square root operation,
which needs an extra Newton-Raphson iteration in double precision.
An issue specific to CPU SIMD kernels
is that LJ pair parameter look-up is costly, as one SIMD load operation
is required for each particle pair. With geometric or Lorentz-Berthelot
combination rules only two loads are required per cluster pair,
latency of which can be hidden with computation.

The CUDA kernels turn out to be instruction latency limited, not memory limited,
although this requires some tricks and a tight packing of the pair list
in memory.
The pseudocode of the kernel is shown in \figref{code_kernel_gpu}. The inner loop calculates interactions
between clusters of $M$=8 and $N$=4, this way 32 threads of a warp calculate
a pair interaction of an entire cluster pair simultaneously.
We chose $N$ smaller than $M$ such that we have more computation
per memory operation. We group cluster pairs two by two in the pair list,
hence the pair search can be done on clusters of 8 particles and the computation
on two 8$\times$4 clusters independently on two warps.
Additionally, we store clusters in range for 8 $i$-clusters in a single pair list with this further improving the data reuse. A $j$-cluster interacts with half
of these 8 $i$-clusters on average, which additionally reduces the memory
pressure by a factor of 4. As the pseudocode on \figref{code_kernel_gpu} shows,
non-interacting $j$-clusters are skipped based on an bitmask-encoded cluster interaction mask,
which only causes a minor overhead.
This hierarchical grouping requires minor modifications in the pair-search code,
only storing the packed exclusions masks becomes more complex.
With this setup we can load the coordinates, atom types, and charges
for 64 $i$-particles in registers on the GPU and thereby maximize
the number of calculations per $j$-particle load. Two warps in a CUDA thread block operate
on a group of 8 $i$-clusters and their $j$-cluster neighbors.
As the two warps by definition access different $j$-particles,
they can run independently and no synchronization is required during computing.
On the Fermi architecture partial forces are accumulated in registers and
reduced in shared memory. In contrast, the Kepler architecture provides a
special ``warp-shuffle'' operation which can be used for efficient synchronization-free
warp-level reduction. 
After a lot of testing and optimization, the CUDA kernels turned
out to be compact and more readable than the CPU SIMD kernels. More code is required for managing GPU device initialization,
kernel launches and transfers between CPU and GPU.

\begin{figure*}
{\small
\begin{verbatim}
/* Each of the MxN i-j pairs is assigned to a thread.
   The sci i-supercluster consists of 8 ci clusters. */
sci = thread block index
for each ci in sci load i-atom data into shared mem.

/* loop over all cj in range of any ci in sci */
for each cj cluster
  load j-i cluster interaction and exclusion mask /* per warp */
  if cj not masked /* non-interacting cj-sci */
    load j-atom data

    /* loop over the 8 i-clusters */
    for each ci cluster in sci
      if cj not masked /* non-interacting cj-ci */
        load i atom data from shared mem.
        r2 = sqrt(|xj –- xi|)
        extract excl_bit exclusion/interaction bit for j-i pair
        if ((r2 < rc_squared) * excl_bit)
          calculate i-j Coulomb and LJ forces
          accumulate i- and j-forces in registers

      reduce j-forces

reduce i-forces
\end{verbatim}
}
\caption{Pseudocode for cluster-pair interaction kernel, NVIDIA GPUs-specific implementation.
\label{code_kernel_gpu}}
\end{figure*}

\subsection{Pair interaction functions}

Calculating energies is only required infrequently in molecular dynamics,
therefore we will concentrate on force-only kernels.
The Lennard-Jones force is:
\begin{equation}
\mathbf{F}_{ij}(\mathbf{r}_{ij}) = e_{ij}\, \left( \frac{12\,C^{12}_{ij}}{ \| \mathbf{r}_{ij} \|^{13}} - \frac{6\,C^{6}_{ij}}{\| \mathbf{r}_{ij} \|^{7}} \right) \frac{\mathbf{r}_{ij}}{\| \mathbf{r}_{ij} \|}
\end{equation}
where $e_{ij}$ is 0 for excluded particle pairs and 1 otherwise.
The LJ coefficients $C^{12}_{ij}$ and $C^{6}_{ij}$ can be different for each atom pair $i$-$j$. In practice there is a limited number of atom types and often combination rules are used to obtain the parameters between two atom types. With x86 SIMD instructions loading arbitrary pair parameters can be costly due to the many load and shuffle operations required.
Using combination rules, either geometric or Lorentz-Berthelot, is more efficient.

The electrostatic interaction form we consider is:
\begin{equation}
\mathbf{F}_{ij}(\mathbf{r}_{ij}) = \frac{q_i q_j}{4\pi \epsilon_0}\left(\frac{e_{ij}}{\| \mathbf{r}_{ij} \|^{2}} + g(\| \mathbf{r}_{ij} \|) \right) \frac{\mathbf{r}_{ij}}{\| \mathbf{r}_{ij} \|}
\end{equation}
Where $g(r)$ is the long-range correction force.
For reaction-field electrostatic we have $g(r)=-2\,k\,r$ with $k$ a constant. This can be evaluated efficiently analytically. For PME we have $g(r)=d\, \mathrm{erf}(\beta\,r)\,r^{-1} / d\,r$, with $\beta$ a constant. Evaluation of the PME correction force is more costly. But as it is bounded and very smooth, linear table interpolation can reach full single precision with a limited table size. As a second option we consider an analytical approximation using a quotient of two polynomials. This requires 24 multiplications and additions and one division to reach full single precision.
Fused multiply-add (FMA) instructions, currently available on GPUs and the AMD Bulldozer
microarchitecture, can speed up this polynomial evaluation significantly.

\section{Performance in practice}
Achieving good performance of load and store intensive kernels requires detailed
understanding of many low-level software optimization aspects: SIMD instruction set,
throughput and latency of instructions on different processor microarchitectures,
cache behavior, as well as experience with compiler-related performance issues.
Unfortunately, it takes a lot of time and effort to reach optimal performance.
Fortunately, this effort is required infrequently and our results can be used by anyone,
as our compute-kernels are released as part of an open source project,
freely available for anyone to use.

\subsection{Hardware and compilers}
To compare the different variants of the algorithm, we focus on the Intel Sandy Bridge CPU architecture.
The reason for this is that at the time of writing this architecture supports AVX, the newest and widest SIMD instruction
set on the x86 platform, and it also provides 256-bit operations. 
This allows a direct comparison between the 4$\times$4 and 4$\times$8 setup, as well as between 128- and 256-bit AVX.
For comparison with other architectures, we also show results on the AMD Bulldozer architecture using 128-bit AVX and FMA instructions as well as NVIDIA Fermi (GF100) and Kepler2 (GK110) \cite{NVIDIA2012} GPUs.
As all CPU architectures in focus support the AVX instruction set, form here on 128- or 256-bit SIMD
will refer to AVX instructions of the respective type. We report all performance
data in cycles which depends only on the microarchitecture, but not the exact CPU or GPU model\footnote{The amount of cache can vary among CPUs with the same microarchitecture (e.g. desktop and server Sandy Bridge), but this does not affect our benchmarks.}.
All CPU kernels were compiled with the GNU C compiler version 4.7.1 with \texttt{-O3}
as the only optimization. Other optimization-related compiler options did not improve the
non-bonded kernel performance. Both Intel C compiler versions 12.1.3 and 13.1.1 produced slightly slower code even with CPU architecture-specific optimizations enabled. The GNU C compiler has greatly improved with recent versions, the difference between version 4.5.3 and 4.7.1 on Sandy Bridge with analytical Ewald kernels is 22\% while with recent Intel compilers even slight regressions have been observed.
The CUDA GPU kernel were compiled with the CUDA compiler version 5.0.7\footnote{
This pre-release version of the 5.0 compiler was used because the final version generates slower code.}
with \texttt{-use\_fast\_math} as well as the architecture-specific optimization
 options \texttt{-arch=sm\_20} and \texttt{-arch=sm\_30}\footnote{Note that with the 5.0 CUDA compilers \texttt{-arch=sm\_35} optimization yields lower performance on GK110 than the \texttt{sm\_30} optimized PTX code JIT compiled by the driver for \texttt{sm\_35}.} for Fermi and Kepler2 GPUs, respectively.

On the CPU we store the properties of the $M$ i-particles in SIMD registers and loop over
the list of clusters of $N$ j-particles. The pair interactions for the $M$ different i-particles are not
interdependent, except that we want to load and store the j-particle properties only once.
There are several choices to be made when transforming this algorithm into actual code. For instruction
(re-)scheduling it is advantageous to write out the $M$ operations for the i-particles,
so it is clear to the compiler that it can reorder them.
For most kernels $N$ matches the SIMD width, but for the 256-bit flavor we also consider
$N$=4, which is half the SIMD width.
On new architectures with wider SIMD units, such as Intel MIC with 16-way SIMD
in single precision,
having $N$ smaller than the SIMD width is even more important.

\subsection{Kernel performance}
Performance of the most important flavors of the fully optimized kernel versions is reported
in \tabref{kernperf_rf} and \tabref{kernperf_ewald} for RF and Ewald, respectively.
The metrics shown in \tabref{kernperf_rf} and \tabref{kernperf_ewald} represent
the peak performance of the respective kernels. The performance of CPU kernels is constant in the regime of 100-100000 particles.
In contrast, GPUs are massively parallel multi-processors which require
a high level of data-parallelism and hence many particle-pairs to reach peak performance.
The CUDA kernels are within 5\% of the peak performance from around 20000 particles; the scaling depends
both on generation of architecture and number of multiprocessors.

We present four different performance metrics.
The the number of pairs calculated per 1000 compute cycles (pairs/cycle)
is the only relevant measure for the raw performance of the algorithm.
The instructions per cycle (IPC) provides an estimate of the hardware utilization.
The last two are the number of floating point operations per pair (flops/pair) and per cycle (flops/cycle), where we try to minimize the former and maximize the latter.
As a reference we show performance for 1$\times$1 kernels which fill the SIMD
unit by unrolling the inner loop over $j$. These kernels do not use LJ
combination rules, as parameters need to be looked up either way, which saves
two floating point operations per pair.
This standard way of employing SIMD results in low performance and low flop
rates (the theoretical peak rate for Intel Sandy Bridge is 8 for 128-bit and
16 for 256-bit instructions, respectively). The high measured IPC indicates that the
instructions are scheduled very efficiently. However, a large part of the instructions load, store
and shuffle data, rather than doing computation.
The 256-bit RF kernel is only 13\% faster than the 128-bit variant while it has
similar IPC. As both kernels execute the same arithmetic instructions,
the observed rather small performance increase is explained by the overhead of
shuffle and data load operations. We aim to address these bottlenecks with
the proposed algorithms by reducing the need for shuffles and loads.
In comparison to the work of Pennycook \cite{Pennycook2013}, here, the effect is much more
pronounced as we need to load two LJ parameters and a charge per $j$ particle, while
they only implement an LJ potential with fixed particle type.
The large drop in performance when using a single thread shows that the 1$\times$1 kernels
are mainly limited by instruction scheduling and HyperThreading (HT)
improves performance by offering the possibility of scheduling instructions
from both threads running on the same physical core.

In double precision with 128-bit AVX we can use 2-way SIMD
and we can compare the performance for small $M$ and $N$.
The 4$\times$2 RF kernel is 26\% faster than the 2$\times$2 kernel,
which outweighs the negative impact of zero interactions in most cases.
Additionally, the pair search for 2$\times$2 takes significant time.
This shows that $M=2$ is not a viable option and we therefore only consider $M=4$ or larger. In the 256-bit kernels we can use the 4$\times$4 scheme
which gives 50\% higher performance than 4$\times$2
and even more on a single thread.

We continue with the single precision kernels for different functional forms.
With 256-bit AVX the $M \times N$ RF kernels have a 3.3 times higher pair rate than 1$\times$1, for Ewald this factor is 2.2. This shows that our approach works.
256-bit is 25\% to 65\% faster than 128-bit depending on the interaction type
and the of use HT.
The performance of the analytical Ewald kernels is similar to that of the tabulated
version with HT, even though the flop rate is very different.
Without HT the tabulated kernels get significantly slower
because of the latencies involved in reading table entries.
The AMD Bulldozer, in contrast with the simultaneous multi-threading Intel HT implements, uses a cluster multi-threading architecture with much of the functional units, including SIMD units, shared between a pair of cores organized in a so called module.
Therefore, we compare performance of a hyperthreaded core on Intel with a module on AMD, both of which support two threads.
Even though Bulldozer has double the theoretical throughput of 4-way SIMD instructions and FMA gives another doubling of the theoretical flop rate, the performance is only marginally higher than 128-bit SIMD on Sandy Bridge.
Moreover, Sandy Bridge using 256-bit SIMD provides a 20\% higher pair rate.
The CUDA GPU kernels provide significantly higher performance when comparing
one streaming multiprocessor with one CPU core. The analytical and tabulated Ewald kernels have similar performance, the former being slightly faster on the Kepler architecture even though this kernel executes about 10\% more instructions. This is explained by the fact that the additional instructions are mainly FMA-s and intrinsics which allow higher instruction level parallelism, higher IPC and better absolute performance than the texture-based table loads.
The analytical PME kernels achieve about half of the real-world peak flop rate,
which is mainly because they don't contain enough FMA instructions.
Also, the presence of conditionals for checking the interaction of each of
the 8 i-clusters in a super-cluster deteriorates the performance by 15\%.

\begin{table*}
\begin{tabular}{ccccccccc}
precision & SIMD width & $M\times N$ & pairs/kcycle & 1 thread & IPC &
flops/pair & flops/cycle \\
\hline
single & 4 & 1$\times$1 &  67 & $-$23\% & 2.32 & 38 & 2.6 \\
single & 8 & 1$\times$1 &  76 & $-$24\% & 2.16 & 38 & 2.9 \\
\\
single & 4 & 4$\times$4 & 175 & $-$19\% & 2.36 & 40 & 7.0 \\
single & 8 & 4$\times$4 & 223 & $-$27\% & 1.96 & 40 & 8.9 \\
single & 8 & 4$\times$8 & 248 & $-$2\% & 1.68 & 40 & 9.9 \\
\\
double& 4 & 2$\times$2 &  52 & $-$30\% & 1.74 & 45 & 2.3 \\
double& 4 & 4$\times$2 &  66 & $-$25\% & 2.16 & 45 & 3.0 \\
double& 8 & 4$\times$4 &  98 & $-$10\% & 1.58 & 45 & 4.4 \\
\end{tabular}
\caption{Performance of the reaction-field SIMD kernels on Intel Sandy Bridge CPU with HT enabled, valid at ``normal'' atomistic simulation conditions with geometric LJ combination rules;
to fully utilize the HT capability of the CPU we use two threads on a single physical core, the performance difference with a single thread per core  (without disabling HT) is also shown. \label{kernperf_rf}}
\end{table*}

\begin{table*}
\begin{tabular}{cccccccccc}
 PU & SIMD width & Ewald & $M\times N$ & pairs/kcycle & 1 thread & IPC &
flops/pair & flops/cycle \\
\hline
 SB & 4 & ana. & 1$\times$1 &  51 & $-$31\% & 2.20 & 66 & 3.4 \\
 SB & 8 & ana. & 1$\times$1 &  63 & $-$16\% & 1.98 & 66 & 4.2 \\
\\
 SB & 4 & tab. & 4$\times$4 & 111 & $-$15\% & 2.42 & 43 & 4.8 \\
 SB & 8 & tab. & 4$\times$4 & 147 & $-$26\% & 2.26 & 43 & 6.3 \\
 SB & 8 & tab. & 4$\times$8 & 134 & $-$6\% & 1.88 & 43 & 5.8 \\
 SB & 4 & ana. & 4$\times$4 & 110 & $-$14\% & 2.40 & 68 & 7.5 \\
 SB & 8 & ana. & 4$\times$4 & 139 & $-$11\% & 1.76 & 68 & 9.5 \\
 SB & 8 & ana. & 4$\times$8 & 137 & +1\% & 1.52 & 68 & 9.3 \\
\\
 BD & 4 & ana. & 4$\times$4 & 114 & $-$23\% & 2.16 & 68 & 7.8 \\
\\
 Fermi   & 32 & tab. & 8$\times$4 & 549 & & 1.66  & 41 & 24   \\
 Kepler2 & 32 & tab. & 8$\times$4 & 1130 & & 3.2 & 41 & 46   \\
 Kepler2 & 32 & ana. & 8$\times$4 & 1151 & & 3.7 & 69 & 85   \\
 
\end{tabular}
\caption{Performance of the single precision Ewald tabulated and analytical SIMD kernels, valid at ``normal'' atomistic simulation conditions with geometric LJ combination rules. PU is the type of SIMD processing unit, SB: one physical Intel Sandy Bridge core (with HT), BD: one AMD Bulldozer module (two cores), and one streaming multiprocessor on NVIDIA Fermi (GF100) and Kepler2 (GK110) GPUs.
On CPUs we consider for comparison an SB core with HT enabled and a BD module, both of which support two threads. Hence, we present performance using two threads; for SB the performance difference when running one thread per core (without disabling HT) is also shown. \label{kernperf_ewald}}
\end{table*}

\section{Effective pair list buffering}

The number of pair interactions calculated in a cycle
reflects how a non-bonded algorithm performs on a certain hardware.
However, this measure is not the best indicator of the effective performance,
since both the buffer region and
the cluster-pair scheme add interactions beyond the cut-off, which by definition evaluate to zero.
To get a useful pair interaction rate which reflects the absolute performance,
only the non-zero interactions should be considered.
As shown on \figref{zeros}, at a commonly used cut-off distance of 1 nm
the 4$\times$4 setup adds 86\% additional pair interactions. However, as shown later, 
with a more than doubled pair interaction evaluation rate, the 4$\times$4 still proves to be faster than 1$\times$1.

\begin{figure}
\centerline{\includegraphics[width=7.5cm]{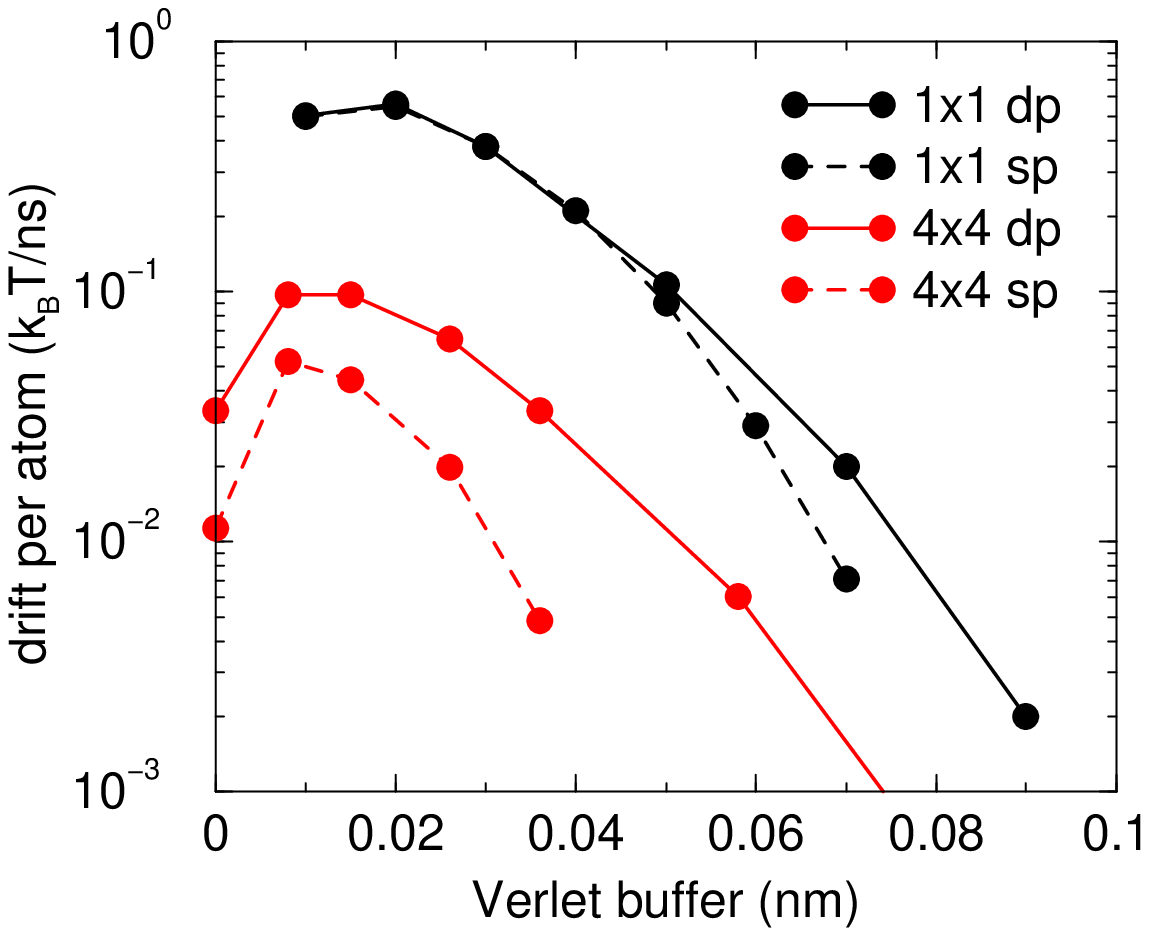}}
\caption{Energy drift per atom in SPC/E water at 298 K as a function of the Verlet buffer size, ''sp`` represents single precision, ''dp`` double precision. A 2 fs time step was used with a pair list update frequency of 10 steps. PME was used with a relative error of 10$^{-5}$ at a cut-off distance of 0.9 nm. The SETTLE algorithm causes negative drift in single precision for large buffers.
\label{drift}}
\end{figure}

Moreover, in certain cases we can actually make use of the extra pairs that the cluster scheme $M \times N $ kernels add.
In molecular dynamics the standard procedure is to ensure that no interacting pair of particles is ever missed. This is usually done by generating a pair list with a so-called Verlet buffer \cite{Verlet1967} which allows particles to move over a small distance without invalidating the pair list. At any step, if any of the particles has moved by more than half the buffer length, the pair list is regenerated \cite{Plimpton1995}.
This condition is sufficient, but not necessary. The pair list needs to be invalidated only
when the distance between a pair of particles decreases by the buffer length, which
will happen far less frequently.
Additionally, this commonly used setup is inconvenient for parallel simulations.
In practice, we can often tolerate small imperfections in the pair list.
In a constant temperature ensemble perfect energy conservation is not a requirement,
as a thermostat will remove excess heat.
Moreover, the amount of energy drift that can be tolerated is very problem dependent.
As there are multiple factors affecting the energy conservation
in a simulation, we can allow the non-bonded interactions to cause a drift of
similar magnitude like all other factors.
If the buffer is too small, some particle pairs which are not in the pair list
can move within the cut-off. We can determine an upper bound to the drift caused by such events
in a constant temperature ensemble, this is derived in the appendix.
The upper bound can be used to set the buffer size for simulations.
With PME, the pair potential
at the cut-off is very small, hence the effect of missing pairs will also be very small.
To quantify this effect, we show the drift as a function of the Verlet buffer size
for SPC/E water \cite{Berendsen87} with a pair list lifetime of 18 fs, see \figref{drift}. This is a representative system as
hydrogens in water are the fastest moving particles in nearly all atomistic simulations. With single precision floating point coordinates, the SETTLE \cite{Miyamoto92} and SHAKE \cite{Ryckaert77} constraint algorithms
cause an energy drift of -0.01 and 0.1 $k_B T$/ns per atom, respectively\footnote{
Note that for SHAKE one should pay attention to how the velocity is corrected \cite{Shaw2007b}}.
The 4$\times$4 setup shows a drift of similar magnitude even without any additional buffering. Thus, in practice, no explicit buffer is required in single precision. One thing to note is that at longer buffer length only repulsive hydrogen pairs contribute to the drift. At zero length, attractive oxygen-hydrogen pairs also contribute which leads to a cancellation of errors.

The effective performance is given by the number of interactions
within the cut-off radius that can be calculated per cycle.
To compare the traditional and cluster schemes we show the performance of 
1$\times$1, and 4$\times$4 256-bit AVX kernels, as well as 8$\times$4 CUDA GPU kernels
with both RF and Ewald electrostatics in \tabref{effperf}.
There is one factor that complicates the comparison. The ratio of the cost
of the search and the force calculation affects the optimal list update
frequency, which in turn affects the required buffer size.
In our implementation, the pair list construction for the 1$\times$1 setup
takes four times longer than calculating the interactions once, where for
a 4$\times$4 setup both take about equal time. We think there is some room
for speed-up in our 1$\times$1 search implementation, which has not been fully optimized.
If we assume we can get it twice as fast, the optimal list update frequency
is somewhere between 10 and 15 steps. The optimal update frequency
for 4$\times$4 and 8$\times$4 is around 10 steps.
For the following comparison we will
use the same update frequency of 10 steps for all setups to simplify
the comparison.
The effective speed-up of the force calculation of the 4$\times$4 over the
1$\times$1 scheme on CPUs is a factor of 1.8 and 1.4 for RF and Ewald electrostatics, respectively.
This speedup is mainly due to higher achieved pair rate, but the smaller buffer also contributes.
Assuming the 1$\times$1 search cost can be brought down to twice the force
calculation cost, the total performance improvement including the search cost
is a factor of 2.0 and 1.5 for RF and Ewald electrostatics, respectively.
The 8$\times$4 scheme results in a lower algorithmic work-efficiency due to the
increase in the ratio of zero interactions calculated.
Note that these results are for a cut-off of 1 nm or 210 non-zero pairs per particle.
With increasing cut-off radius, the efficiency increases and the performance
improvement approaches a factor of 3.
As we run the pair search on the, slower, CPU, a longer list update interval
often provides better total performance. The GPU kernels use a conditional
for skipping pairs beyond the cut-off, unlike the CPU $M \times N$ kernels,
which use masking.
Therefore the pair-rate increases with buffer size. But calculating more pair
distances always decreases the effective performance.
Still, the cluster algorithm demonstrates the potential of streaming architectures with an effective performance of a factor of 5 and 7 higher than the 4$\times$4 CPU RF and Ewald kernels, respectively.

\begin{table*}
\begin{tabular}{ccccccccccccc}
PU & elec. & $M \times N$ & pairs/kcycle & list upd.  & buffer & effective  & ratio & effective \\
   &       &              &              & (steps) &  (nm)  & pair ratio & vs 1$\times$1 & pairs/kcycle \\
\hline
SB      & RF    & 1$\times$1 &   76 & 10 & 0.09 & 0.77 &      &  59 \\
SB      & RF    & 4$\times$4 &  223 & 10 & 0.07 & 0.48 & 0.62 & 107 \\
Kepler2 & RF    & 8$\times$4 & 1351 & 10 & 0.07 & 0.40 & 0.52 & 544  \\
Kepler2 & RF    & 8$\times$4 & 1386 & 20 & 0.10 & 0.37 & 0.48 & 519  \\
SB      & Ewald & 1$\times$1 &   63 & 10 & 0.05 & 0.86 &      &  54 \\
SB      & Ewald & 4$\times$4 &  139 & 10 & 0.00 & 0.53 & 0.61 &  74 \\
Kepler2 & Ewald & 8$\times$4 & 1151 & 10 & 0.00 & 0.47 & 0.55 & 545 \\
Kepler2 & Ewald & 8$\times$4 & 1181 & 20 & 0.03 & 0.44 & 0.51 & 521 \\
\end{tabular}
\caption{Effective performance of the 256-bit AVX and Keper2 CUDA kernels in single precision at an energy drift of 0.1 $k_B T$/ns per atom for water at 298 K with a 1 nm cut-off and a 2 fs time step.
The last three columns show the ratio of non-zero and total calculated pairs 
and the effective performance, only counting the non-zero pairs.
The second-last column shows the effective ratio of $M \times N$ versus
$1 \times 1$ with a pair-list update every 10 steps.
\label{effperf}}
\end{table*}

\section{Conclusions}
For calculating non-bonded interactions in molecular simulations,
the standard particle-based pair interaction algorithms
commonly SIMD-parallelized by loop unrolling have reached their limits.
Kernels based on these approaches are often limited by the high memory to arithmetic
operation ratio, the number of data shuffle operations required,
and restrictions in instruction scheduling which reduces the potential for memory latency hiding.
The pair list construction is affected by
the same issues. We have presented a simple and flexible approach
to overcome these problems. A scheme using cluster pairs of $M$ versus $N$ atoms
leads to kernels that efficiently utilize current CPU and GPU SIMD units.
The memory pressure is reduced by a factor $M$ on CPUs. We found that $M$=4
usually provides the best performance. On GPUs we use $M$=8 and
the memory pressure is reduced by another factor of 4 by loading
and operating on up to 8 $i$-clusters at once.
The algorithm reorganizes the data representation at the lowest, particle level. Therefore, any method in
the literature that applies to particles, can be applied to the clusters
in our method. An example used here is the Verlet buffer. While the widely
used linked cell list for reducing the search space can be applied,
this does not offer any advantage as the locality of the clusters is already
available through the grid used to generate the clusters.
The performance advantage of our method over traditional algorithms depends
on the computational cost of the interactions, the number of particle pairs
within the cut-off and the SIMD width. While in many cases our cluster-based
algorithm significantly outperforms the particle-based algorithms, in some
cases it can be less advantageous. For cheap interactions the
reduction of shuffling and memory operations will favor the cluster setup,
whereas for expensive interactions the extra
zero interactions can outweigh the gains.

For typical atomistic molecular simulations our method performs
very well and is a factor 1.5 to 3 faster on 8-wide SIMD than traditional methods.
On Intel Sandy Bridge CPUs as well as CUDA GPUs the flop rate is above 60\%
of the peak.
Most importantly, our scheme inherently maps well to future CPU and GPU
architectures as well as existing ones not discussed here.
As the number of floating
point operation per load/store operation can be tuned, a reduction of the
arithmetic cycles per kernel, e.g. by introduction of FMA instructions, will result
in higher performance. Additionally, wider SIMD units, for example 16-way SIMD in Intel
Xeon Phi, can be used efficiently with a limited amount of effort.

\section*{Funding}
This work was supported by the European Research Council
(grants nr. 258980 and nr. 209825), the Swedish e-Science Research Center
and the ScalaLife EU FP7 project.

\section*{Acknowledgments}
The authors thank Erik Lindahl for providing the analytical approximation of
the Ewald correction force and for his advice on x86 SIMD optimization,
NVIDIA for advice on CUDA optimization and Mark Abraham for thoroughly
reviewing the code and this manuscript.

\section*{Appendix}
For a canonical ensemble, an upper bound on the average energy drift
due to the finite Verlet buffer size can be derived. This depends on
the atomic displacements and the shape of the potential at the cut-off.
The displacement-distribution along one dimension for a freely moving particle
with mass $m$ over time $t$ at temperature $T$ is Gaussian with zero mean
and variance $\sigma^2 = t\,k_B T/m$.
The variance of the distance between two non-interacting particles is $\sigma^2 = \sigma_{12}^2 = t\,k_B T(1/m_1+1/m_2)$.
In practice, particles interact with each other over time $t$.
These interactions make the displacement distribution narrower,
since any interaction will hinder free motion of particles.
Ignoring the effect of interactions on the displacements thus provides an upper bound.
We calculate interactions with a non-bonded interaction cut-off distance
of $r_c$ and a pair list cut-off of $r_\ell=r_c+r_b$, where $r_b$ is
the Verlet buffer size.
We can then write the average energy drift over time $t$ for pair
interactions between a particle of type 1 surrounded by particles
of type 2 with number density $\rho_2$, when the inter-particle distance
changes from $r_0$ to $r_t$, as:
\lipsum[1]
\begin{widetext}
\begin{eqnarray}
\langle \Delta V \rangle \! &=&
\int_{0}^{r_c} \int_{r_\ell}^\infty 4 \pi r_0^2 \rho_2 V(r_t) G\!\left(\frac{r_t-r_0}{\sigma}\right) d r_0\, d r_t \\
&\approx&
\int_{-\infty}^{r_c} \int_{r_\ell}^\infty 4 \pi r_0^2 \rho_2 \Big[ V'(r_c) (r_t - r_c) +
\\
& &
\phantom{\int_{-\infty}^{r_c} \int_{r_\ell}^\infty 4 \pi r_0^2 \rho_2 \Big[}
 V''(r_c)\frac{1}{2}(r_t - r_c)^2 \Big] G\!\left(\frac{r_t-r_0}{\sigma}\right) d r_0 \, d r_t\\
&\approx&
4 \pi (r_\ell+\sigma)^2 \rho_2
\int_{-\infty}^{r_c} \int_{r_\ell}^\infty \Big[ V'(r_c) (r_t - r_c) +
\\
& &
\phantom{4 \pi (r_\ell+\sigma)^2 \rho_2 \int_{-\infty}^{r_c} \int_{r_\ell}^\infty \Big[}
V''(r_c)\frac{1}{2}(r_t - r_c)^2 \Big] G\!\left(\frac{r_t-r_0}{\sigma}\right) d r_0 \, d r_t\\
&=&
4 \pi (r_\ell+\sigma)^2 \rho_2 \bigg\{
\frac{1}{2}V'(r_c)\left[r_b \sigma G\!\left(\frac{r_b}{\sigma}\right) - (r_b^2+\sigma^2)E\!\left(\frac{r_b}{\sigma}\right) \right] +
\\
& &
\phantom{4 \pi (r_\ell+\sigma)^2 \rho_2 \bigg\{ }
\frac{1}{6}V''(r_c)\left[ \sigma(r_b^2+\sigma^2)G\!\left(\frac{r_b}{\sigma}\right) - r_b(r_b^2+3\sigma^2 ) E\!\left(\frac{r_b}{\sigma}\right) \right]
\bigg\}.
\end{eqnarray}
\end{widetext}
\lipsum[1]
Here, $G(x)$ is a Gaussian distribution with zero mean, unit variance, and $E(x)=\frac{1}{2}\mathrm{erfc}(x/\sqrt{2})$. We always want to achieve small energy drift, so $\sigma$ will be small compared to both $r_c$ and $r_\ell$. Thus, the approximations in the above equations are good since the Gaussian distribution decays rapidly. To calculate the total energy drift, the drift needs to be averaged over all particle pairs and weighted with the particle count.

\bibliographystyle{model1a-num-names}

\end{document}